\documentclass[
    ,final            % use final for the camera ready runs
  ]
  {aipproc}

\layoutstyle{6x9}

\newcommand{\be}{\begin{equation}}
\newcommand{\ee}{\end{equation}}
\newcommand{\bea}{\begin{eqnarray}}
\newcommand{\eea}{\end{eqnarray}}

\begin{document}

\title{Modelling Drell-Yan pair production in $\bar pp$ \footnote{Work supported by BMFT.}}
\author{O.~Linnyk }{
  address={Institut f\"ur Theoretische Physik, Universit\"at Gie\ss en,
  Germany},
  altaddress={olena.linnyk@theo.physik.uni-giessen.de}
}
\author{S.~Leupold}{
  address={Institut f\"ur Theoretische Physik, Universit\"at Gie\ss en,
  Germany}
}
\author{U.~Mosel}{
  address={Institut f\"ur Theoretische Physik, Universit\"at Gie\ss en,
  Germany}
}
\begin{abstract}
We predict the triple differential cross section of the unpolarized Drell-Yan process at $\sqrt{S}=6$~GeV.
The model incorporates primordial parton transverse momentum and quark off-shellness
effects caused by the initial state interaction.
\end{abstract}
\maketitle

%%%%%%%%%%%%%%%%%%%%%%%%%%%%%%%%%%%%%%%%%%%%
%% MAINMATTER
%%%%%%%%%%%%%%%%%%%%%%%%%%%%%%%%%%%%%%%%%%%%

%%%%%%%%%%%%%%%%%%%%%%%% Motivation %%%%%%%%%%%%%%%%%%%%%%%%%%%%%%%%%

The study of the Drell-Yan lepton pair production process ($NN \to \mu ^+ \mu
^- X$) improves our understanding of the quark-gluon  structure
of the nucleon. It provided the most accurate data on the sea quark
distribution in the nucleon and also additionally constrained the
valence quark distributions earlier measured in deep inelastic scattering. 
The PANDA~\cite{PANDA} experiment at the  future GSI facility will use the 
$\bar p p \to \mu ^+ \mu ^- X$ reaction to pin down the polarized quark distributions in
the proton.

The conventional perturbative QCD (pQCD) approach to the calculation of hard
scattering cross sections assumes that the partons in hadrons are collinear and on-shell. 
This is equivalent to using the impulse approximation, {\it i.e.} neglecting
the initial and finite state interactions (ISI and FSI).
Next to leading order pQCD calculations reproduce the Drell-Yan
cross section integrated over the transverse  momentum of the lepton pair 
($p_T$). However, it fails to reproduce the un-integrated triple differential
cross section. A number of effects missed in the  standard perturbative
treatment are essential to explain the observed $p_T$-spectrum of the lepton
pairs. Namely,
\begin{itemize}
\item the interaction of the active quark with the spectators,
\item intrinsic transverse momentum of partons ($k_T$),
\item soft gluon radiation. 
\end{itemize}
Nonperturbative effects are especially sizable at low $\sqrt{S}$ and high
ratios $M^2/S$, where $M$ is the mass of Drell-Yan pair, which is exactly the
kinematical region to be probed by PANDA.
The aforesaid effects can not be described in any fixed order of the perturbative
expansion and require a resummation of an infinite number  of diagrams and
modelling of the non-perturbative higher twist contributions. 

%%%%%%%%%%%%%%%%%%%%%%% Model %%%%%%%%%%%%%%%%%%%%%%%%%%%%%%%%%%%%%%%%%

\begin{figure*}
%\begin{center}
%\subfigure[$M = 1$~GeV] % caption for subfigure a
%{
         \resizebox{0.75\textwidth}{!}{%
         \includegraphics{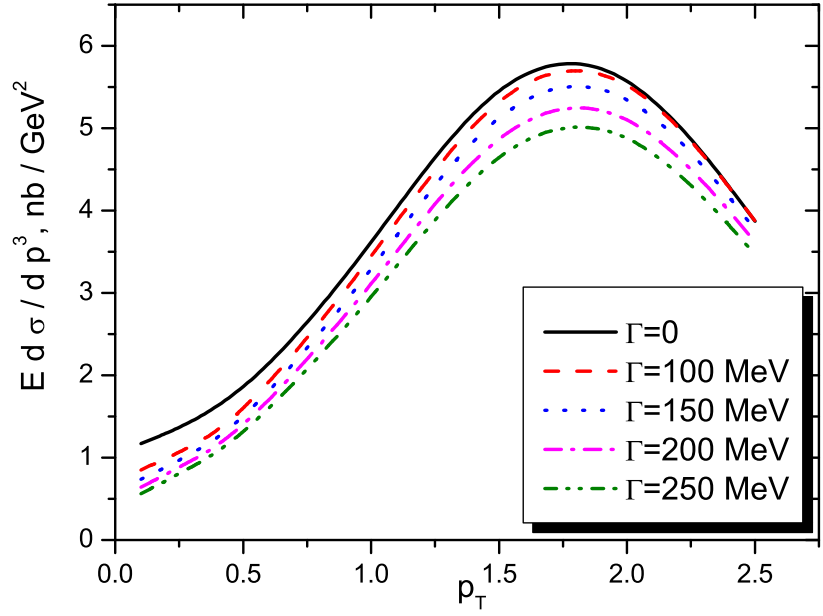}
         } 
%}
\vspace{-0.5 cm}
\caption{
Prediction for the $p_T$-distribution of Drell-Yan lepton pairs in $\bar pp$ collision at $\sqrt{S}=6$~GeV, $M=1$~GeV. 
Avarage quark intrinsic transverse momentum is 1~GeV. 
The solid line is the result of calculations in the intrinsic-$k_T$ approach
(width $\Gamma \!=\! 0$). 
The other curves are generated with $\Gamma$ in the range determined by fitting 
existing Drell-Yan data. 
$x_F\! =\! 0.1$ in all plots.
}
\label{p1} 
\end{figure*}

\begin{figure*}
%\subfigure[$M = 2$~GeV] % caption for subfigure b
%{
         \resizebox{0.75\textwidth}{!}{%
         \includegraphics{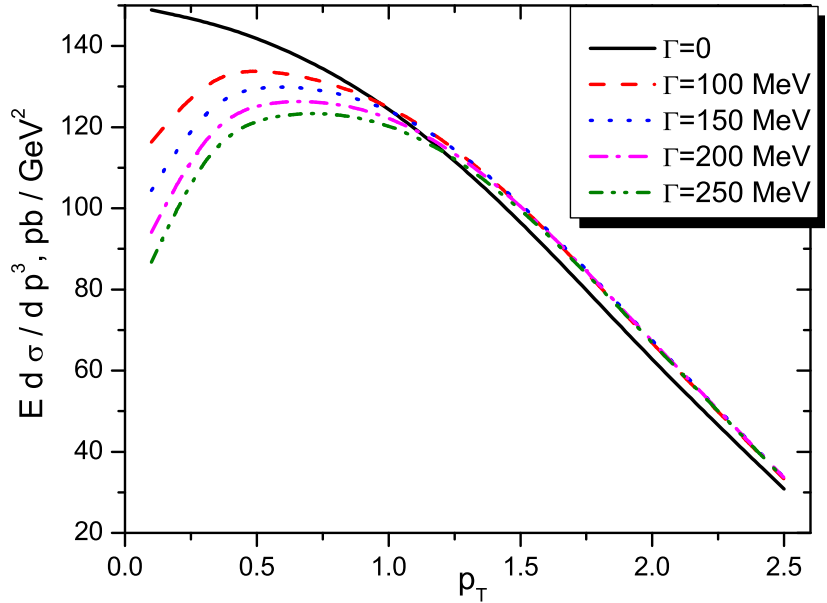}
         } 
%}
%\end{center}
\vspace{-0.5 cm}
\caption{
Same as Fig.~\ref{p1}, but for the Drell-Yan pair mass $M=2$~GeV.
}
\label{p2} 
\end{figure*}

We have parametrized the effects listed above by means of $k_T$-unintegrated parton
distributions and a phenomenological  spectral function for (anti-) quarks in
the nucleon. 
The non-vanishing intrinsic transverse momentum $k_T$ and the off-shellness 
$m^2\!\equiv \!k^+ k^- \! - \! k_T^2$ of the partons are generated by the ISI.
The off-shellness distribution is taken as a Breit-Wigner in $m$ with a constant width $\Gamma$:
\be 
\label{BW}
\mbox{A}(m,\Gamma) = \frac{1}{\pi}\frac{\Gamma}{m^2+\frac{1}{4}\Gamma^2} .
\ee
Additionally, we take into account the exact off-shell kinematics.
The model was developed in~\cite{paper2}, 
where it was applied to calculate the ISI effects in deep inelastic
scattering~(DIS) and the Drell-Yan process.

The calculations in this approach  are in excellent agreement with Drell-Yan 
data from Fermilab at projectile energies of 800~GeV/c and 125~GeV/c~\cite{paper2}. 
The $p_T$-distribution of the lepton pairs  $d\sigma /d M^2 d p_T^2 d x_F$ 
was reproduced  as well as the partly integrated (double-differential) cross sections.
Here, $x_F$ is the Feynman variable of the produced lepton pair. Both shape and
magnitude of the observed triple differential cross sections were reproduced very
well in all the bins of $M$, $x_F$ and $p_T$ without a need for a K-factor.

The success of our description of the data over the wide range of $\sqrt{S}$
allowed us to extrapolate the values of the two model parameters ($k_T$  dispersion and
the quark width) to $\sqrt{S}=6$~GeV and $M < 6$~GeV. 
In this contribution, we present a prediction for the triple  differential Drell-Yan cross
section in the kinematical region of this low $\sqrt{S}$ and high ratio
$M/\sqrt{S}$.

%%%%%%%%%%%%%%%%%%%%%%%% Results %%%%%%%%%%%%%%%%%%%%%%%%%%%%%

\begin{figure*}
%\begin{center}
%\subfigure[$M = 3$~GeV] % caption for subfigure a
%{
         \resizebox{0.75\textwidth}{!}{%
         \includegraphics{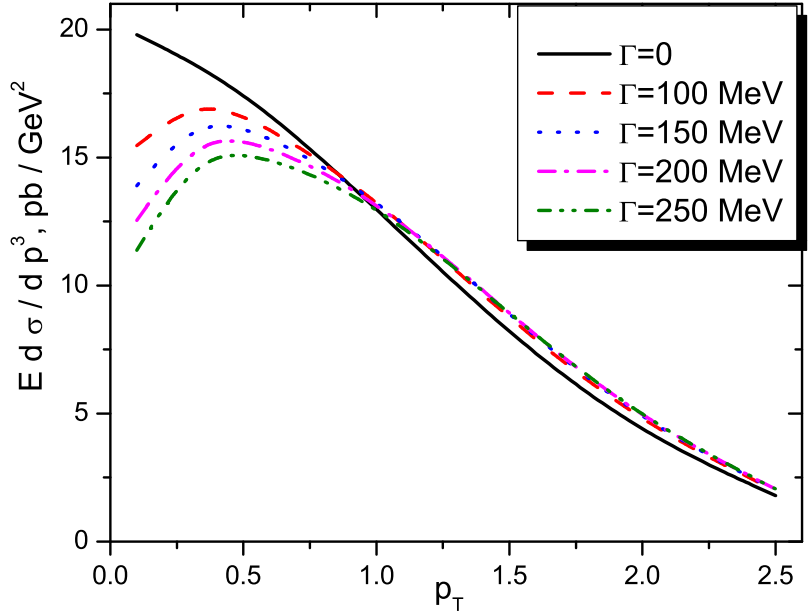}
         } 
%}
%\hspace{-0.5cm}
\vspace{-0.5 cm}
\caption{Same as Fig.~\ref{p1}, but for the Drell-Yan pair mass $M=3$~GeV.
}
\label{p3} 
\end{figure*}

%\section{Results}
\label{results}

\begin{figure*}
%\subfigure[$M = 5$~GeV] % caption for subfigure b
%{
         \resizebox{0.75\textwidth}{!}{%
         \includegraphics{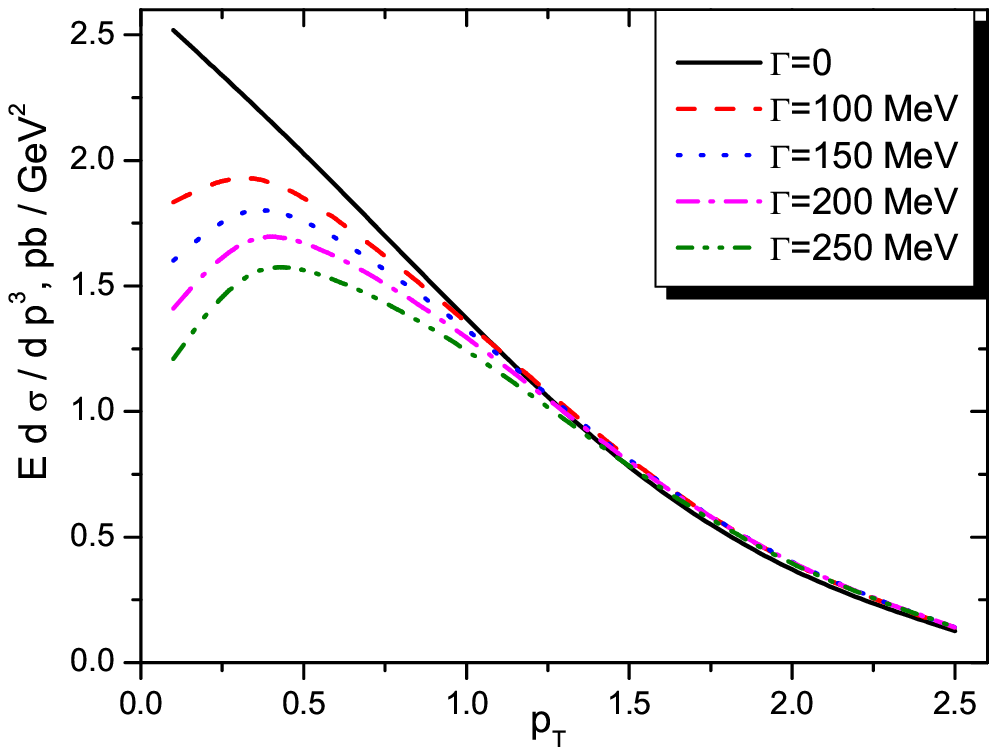}
         } 
%}
%\end{center}
\vspace{-0.5 cm}
\caption{Same as Fig.~\ref{p1}, but for the Drell-Yan pair mass $M=5$~GeV.
}
\label{p4} 
\end{figure*}

The prediction for the transverse momentum distribution of the
Drell-Yan pairs at $\sqrt{S}=6$~GeV is presented in Fig.~\ref{p1}-\ref{p4} for
different masses of the produces pairs.
We have plotted the observable
\be
 \frac{d\sigma}{d\vec p}
\equiv \frac{2}{\pi \sqrt{s}}\frac{d\sigma}{d x_F d p_T ^2}
=\frac{2}{\pi \sqrt{s}} \int \limits _{\mbox{\small bin}} \! 
                                \frac{d\sigma}{d x_F d p_T ^2 dM^2} \, dM^2 
%\mbox{ where}
\label{triple}
\ee
for $M=1,2,3,5$~GeV.
The plot for $M=4$~GeV is not shown, because the Drell-Yan process in this region can not be experimentally disentangled from the charmonium production.
Note that the scale changes from nb in Fig.~\ref{p1} to pb in the other figures.
Note also the qualitative difference of the cross section at the Drell-Yan
pair mass $M=1$~GeV from the other plots. In contrast to the higher mass bins,
the peak of the $p_T$-distribution in Fig.~\ref{p1} is not around zero. This behavior
appears, when $M$ approaches the partonic avarage intrinsic transverse momentum, 
{\it i.e. }in the distribution of the low virtuality
photons, which can be produced by the partonic transverse motion alone. It is
worthwhile to stress that this drastic change in the $p_T$-dependence of the
cross section takes place for all values of $\Gamma$. An experimental
verification of this effect would be a direct test for the transverse
momentum distribution of quarks in the proton. 

We have generated several theoretical curves with  $\Gamma \! = \! (100-250)$~MeV,
which is the range determined from fitting the Drell-Yan data at higher
energies.
The solid lines are the results of our calculations in the intrinsic-$k_T$ approach
($\Gamma \! = \! 0$)~\cite{paper2}.
The uncertainty of $\Gamma$ reflects the lack of data in the low $\sqrt{S}$ region. 
We expect that PANDA data will allow us to constrain
the quark spectral function in the proton with a much better accuracy (see
\cite{paper3} for details).

%%%%%%%%%%%%%%%%%%%%%%%%%%%%%%%%%%%%%%%%%%%%
%% Sample figure:
%%
%% The option [height=...] scales the picture to the given height,
%% without it it would be printed at its nominal size
%%%%%%%%%%%%%%%%%%%%%%%%%%%%%%%%%%%%%%%%%%%%
%\begin{figure}
%  \includegraphics[height=.3\textheight]{golfer}
%  \caption{Picture to fixed height}
%  \caption{Picture to fixed height}
%\end{figure}

%%%%%%%%%%%%%%%%%%%%%%%%%%%%%%%%%%%%%%%%%%%%%%%%
%% BACKMATTER
%%%%%%%%%%%%%%%%%%%%%%%%%%%%%%%%%%%%%%%%%%%%%%%%

%\begin{theacknowledgments}
%  Infandum, regina, iubes renovare dolorem, Troianas ut opes et
%  lamentabile regnum cruerint Danai; quaeque ipse miserrima vidi, et
%  quorum pars magna fui. Quis talia fando Myrmidonum Dolopumve aut duri
%  miles Ulixi temperet a lacrimis?
%\end{theacknowledgments}



\bibliographystyle{aipproc}   % if natbib is available

%\bibliography{sample}

%  \typeout{******************************************}
%  \typeout{** Please run "bibtex \jobname" to optain}
%  \typeout{** the bibliography and then re-run LaTeX}
%  \typeout{** twice to fix the references!}
%  \typeout{******************************************}

%%%%%%%%%%%%%%%%%%%%%%%%%%%%%%%%%%%%%%%%%%%
%% The following lines show an example how to produce a bibliography
%% without the help of the BibTeX program. This could be used instead
%% of the above.
%%%%%%%%%%%%%%%%%%%%%%%%%%%%%%%%%%%%%%%%%%%

\end{document}